\documentclass[twocolumn,english,superscriptaddress,floatfix,showpacs]{revtex4}
\usepackage[T1]{fontenc}
\usepackage[latin9]{inputenc}
\usepackage{amsmath}
\usepackage{amssymb}
\usepackage{graphicx}
\usepackage{esint}

\makeatletter
\@ifundefined{textcolor}{}
{%
 \definecolor{BLACK}{gray}{0}
 \definecolor{WHITE}{gray}{1}
 \definecolor{RED}{rgb}{1,0,0}
 \definecolor{GREEN}{rgb}{0,1,0}
 \definecolor{BLUE}{rgb}{0,0,1}
 \definecolor{CYAN}{cmyk}{1,0,0,0}
 \definecolor{MAGENTA}{cmyk}{0,1,0,0}
 \definecolor{YELLOW}{cmyk}{0,0,1,0}
}

\@ifundefined{definecolor}
 {\usepackage{color}}{}
\usepackage{babel}
\usepackage{babel}
\usepackage{babel}
\usepackage{babel}
\usepackage{babel}
\usepackage{babel}
\usepackage{babel}
\usepackage{babel}
\usepackage{babel}
\usepackage{babel}
\usepackage{babel}
\usepackage{babel}

\@ifundefined{textcolor}{}{
 \definecolor{BLACK}{gray}{0}
 \definecolor{WHITE}{gray}{1}
 \definecolor{RED}{rgb}{1,0,0}
 \definecolor{GREEN}{rgb}{0,1,0}
 \definecolor{BLUE}{rgb}{0,0,1}
 \definecolor{CYAN}{cmyk}{1,0,0,0}
 \definecolor{MAGENTA}{cmyk}{0,1,0,0}
 \definecolor{YELLOW}{cmyk}{0,0,1,0}
}
\@ifundefined{definecolor}{
}{}
\@ifundefined{definecolor}{\usepackage{color}}{}

\makeatother

\usepackage{babel}
\begin{document}

\title{Scaling between magnetic and lattice fluctuations in\textmd{ }iron-pnictide
superconductors}

\author{Rafael M. Fernandes}

\affiliation{School of Physics and Astronomy, University of Minnesota, Minneapolis,
MN 55116, USA}

\author{Anna E. Böhmer}

\affiliation{Institut für Festkörperphysik, Karlsruher Institut für Technologie,
76021 Karlsruhe, Germany}

\author{Christoph Meingast}

\affiliation{Institut für Festkörperphysik, Karlsruher Institut für Technologie,
76021 Karlsruhe, Germany}

\author{Jörg Schmalian}

\affiliation{Institut für Theorie der Kondensierten Materie, Karlsruher Institut
für Technologie, 76128 Karlsruhe, Germany}

\date{\today }
\begin{abstract}
The phase diagram of the iron arsenides is dominated by a magnetic
and a structural phase transition, which need to be suppressed in
order for superconductivity to appear. The proximity between the two
transition temperature lines indicates correlation between these two
phases, whose nature remains unsettled. Here, we find a scaling relation
between nuclear magnetic resonance (NMR) and shear modulus data in
the tetragonal phase of electron-doped $\mathrm{Ba(Fe_{1-x}Co_{x})_{2}As_{2}}$
compounds. Because the former probes the strength of magnetic fluctuations
while the latter is sensitive to orthorhombic fluctuations, our results
provide strong evidence for a magnetically-driven structural transition. 
\end{abstract}


\maketitle
The fact that superconductivity in most iron arsenide materials appears
in close proximity to a magnetic phase transition \cite{reviews}
led to the early proposal that magnetic fluctuations play a fundamental
role in promoting Cooper pairing \cite{reviews_pairing}. Indeed,
the NMR spin-lattice relaxation rate $1/T_{1}$, which is proportional
to the strength of spin fluctuations, is substantially enhanced in
optimally doped compounds, where the superconducting transition temperature
$T_{c}$ acquires its highest value, and rather small in strongly
overdoped samples, where $T_{c}$ vanishes \cite{Ning10}. At the
same time, a tetragonal-to-orthorhombic phase transition is always
found near the magnetic transition, and, consequently, in the vicinity
of the superconducting dome \cite{Fisher11,Fisher12}. Measurements
of the shear elastic modulus $C_{s}$, which is the inverse susceptibility
of the orthorhombic distortion, also found a clear correlation between
the strength of lattice fluctuations and $T_{c}$ \cite{FernandesPRL10,Yoshizawa12,Bohmer}.
Therefore, the road towards the understanding of the high-temperature
superconducting state in the iron pnictides necessarily passes through
the understanding of the relationship between the magnetic and structural
transitions.

\begin{figure}
\begin{centering}
\includegraphics[width=0.85\columnwidth]{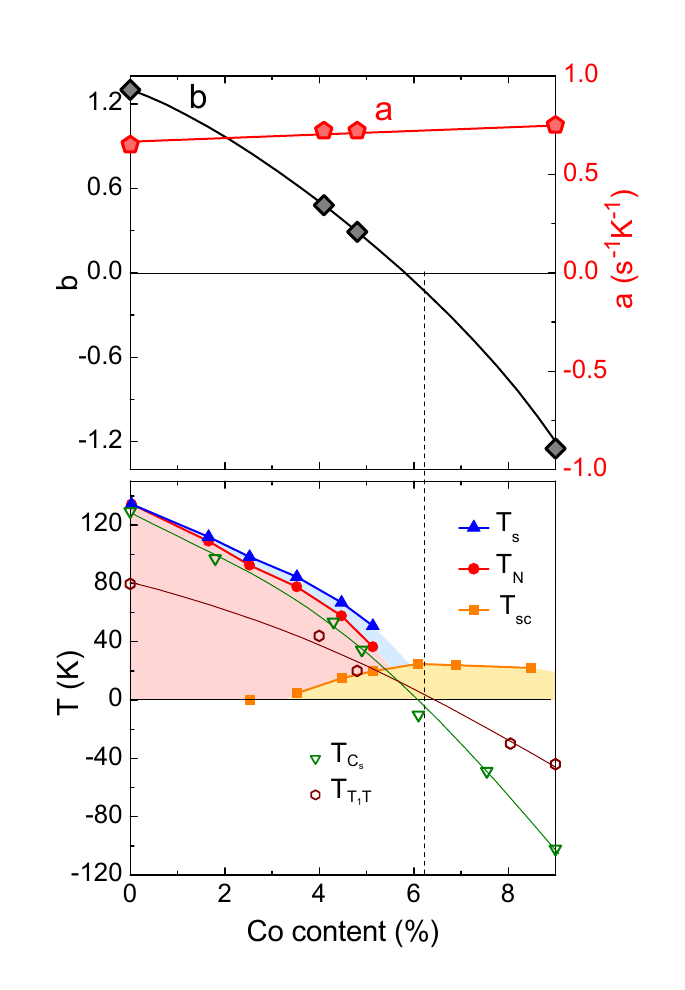} 
\par\end{centering}

\caption{Doping dependence of the fitting parameters $a$ and $b$, and the
phase diagram of $\mathrm{Ba(Fe_{1-x}Co_{x})_{2}As_{2}}$. Notice
that $a$ remains nearly constant, whereas $b$ aproaches zero near
optimal doping, where the superconducting transition temperature is
the highest. The transition temperatures in the phase diagram were
obtained from resistivity measurements of Ref. \cite{Chu09}. The
bare transition temperatures $T_{T1T}$ and $T_{C_{s}}$ are obtained
from Curie-Weiss fittings of the data from Ref. \cite{Ning10} and
\cite{Bohmer}, respectively. Lines are a guides to the eye.}

\label{fig_phase_diagram} 
\end{figure}

That these two phases are correlated is evident from the phase diagram
of the iron pnictides, since the structural transition line (at temperature
$T_{s}$) and the magnetic transition line (at temperature $T_{N}\leq T_{s}$)
follow each other closely in the normal state as doping (or pressure)
is changed \cite{Nandi10,Kim11,Birgeneau11} (see Fig. \ref{fig_phase_diagram}).
Although strong evidence has been given for an electronic mechanism
driving the structural transition \cite{Fisher12}, the key unresolved
issue is its microscopic nature. Two competing approaches have been
proposed, where magnetic fluctuations play a fundamentally distinct
role. One point of view is based on strong inter-orbital interactions
that lead to orbital order and may induce magnetism as a secondary
effect \cite{w_ku10,Phillips10,devereaux10,Phillips12,Stanev13}.
Alternatively, spin fluctuations are considered the driving force
behind the structural transition by inducing strong nematic \cite{Xu08,Fang08,Fernandes12,Indranil11}
or closely related orbital fluctuations \cite{Kontani12,Kontani_solidstate}.
While it is clear that both degrees of freedom are important to correctly
describe the electronic orthorhombic phase \cite{SUST12,Dagotto13},
it is crucial to establish which of the two is the primary one, since
both orbital \cite{KontaniPRB12,Ono10} and spin fluctuations \cite{reviews_pairing}
have been proposed as candidates for the unconventional pairing state
of the pnictides.

\begin{figure}
\begin{centering}
\includegraphics[width=0.85\columnwidth]{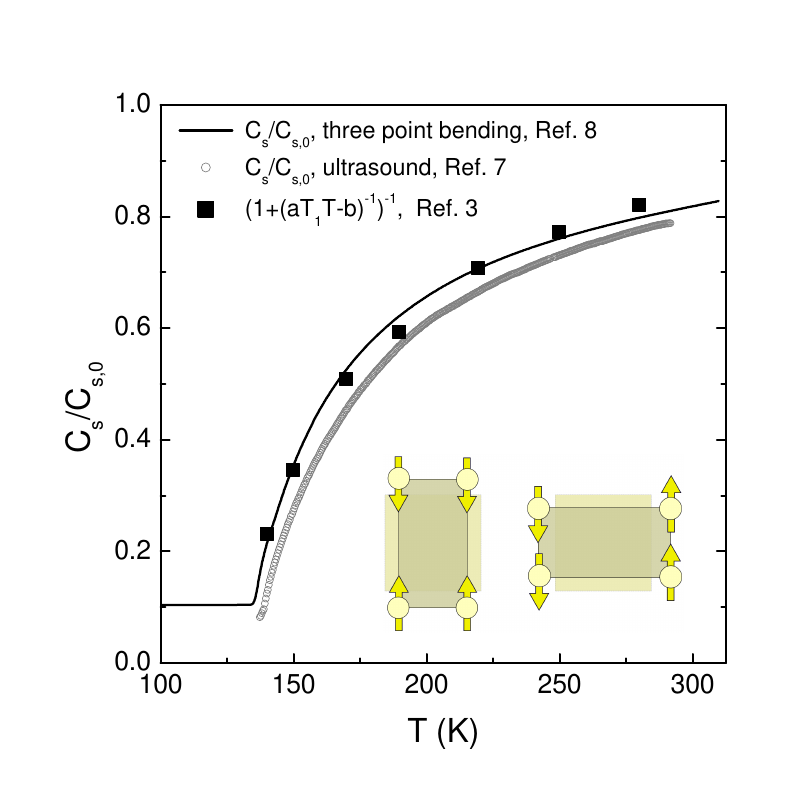} 
\par\end{centering}

\caption{Scaling between the shear modulus $C_{s}$ (open symbols, from Ref.
\cite{Yoshizawa12}, continuous line from Ref. \cite{Bohmer}) and
the NMR spin-lattice relaxation rate $1/T_{1}$ (closed symbols, from
Ref. \cite{Ning10}) in the tetragonal phase of the parent compound
$\mathrm{BaFe_{2}As_{2}}$. $C_{s,0}$ denotes the non-critical, high-temperature
shear modulus. The fitting parameters are $a=0.65$ (sK)$^{-1}$ and
$b=1.3$ in Eq. (\ref{scaling}). The inset schematically represents
the magneto-elastic coupling, which makes bonds connecting anti-parallel
(parallel) spins expand (shrink). }

\label{fig_parent} 
\end{figure}

Differentiating between the two proposed scenarios is difficult in
the symmetry-broken state, because all possible order parameters are
non-zero at $T<T_{s}$: orthorhombicity, orbital polarization, magnetic
anisotropy, etc \cite{Davis10,Duzsa11,Uchida11,ZXshen11,Greene12,Matsuda12,Imai12}.
Instead, additional information can be obtained by studying the fluctuations
associated with each degree of freedom in the tetragonal phase at
$T>T_{s}$ \cite{Gallais13}. In this regime, it holds generally that
the electronic driven softening of the elastic shear modulus $C_{s}$
is determined by the static susceptibility \textbf{$\chi_{\mathrm{nem}}\equiv\left\langle \varphi\varphi\right\rangle $}
associated with the electronic degree of freedom \textbf{$\varphi$}
that drives the structural transition: 
\begin{equation}
\frac{C_{s}}{C_{s,0}}=\left(1+\frac{\lambda^{2}}{C_{s,0}}\chi_{\mathrm{nem}}\right)^{-1}\label{chinem}
\end{equation}

The tetragonal symmetry is broken once \textbf{$\left\langle \varphi\right\rangle \neq0$},
which gives rise to a finite orthorhombic distortion $\epsilon_{xx}-\epsilon_{yy}\propto\left\langle \varphi\right\rangle $.
In the orbital fluctuations model, $\varphi$ corresponds to the difference
between the occupations of the \textbf{$d_{xz}$} and \textbf{$d_{yz}$}
orbitals, and \textbf{$\lambda$} is the coupling between the lattice
and orbital distortions \cite{Phillips10}. On the other hand, in
the spin-nematic case, \textbf{$\varphi$ }is an Ising-type degree
of freedom referring to the relative orientation of neighboring spin
polarizations and $\lambda$ is the magneto-elastic coupling \cite{FernandesPRL10}.
\textbf{$C_{s,0}$} is the bare shear modulus in the absence of these
electronic degrees of freedoms. 

In this paper, we show that the nematic susceptibility \textbf{$\chi_{\mathrm{nem}}$}
is closely related to the dynamic spin susceptibility, strongly supporting
a magnetically-driven structural transition in the \textbf{$\mathrm{Ba(Fe_{1-x}Co_{x})_{2}As_{2}}$}
family of pnictides. Specifically,\textbf{ }we show that spin fluctuations,
given by the NMR spin-lattice relaxation rate $1/T_{1}$, and orthorhombic
fluctuations, given by the shear modulus $C_{s}$, satisfy the scaling
relation:

\begin{equation}
\frac{C_{s}}{C_{s,0}}=\frac{1}{1+\left[a\left(T_{1}T\right)-b\right]^{-1}}\label{scaling}
\end{equation}
with doping-dependent, but temperature-independent constants $a$
and $b$. The spin-lattice relaxation rate and the shear modulus data
considered here are the ones previously presented in Refs. \cite{Ning10}
and \cite{Bohmer}, respectively. The raw data contains contributions
from critical fluctuations - which is our interest here - and non-critical
processes, which are unrelated to the magnetic or structural transitions.
To make a meaningful scaling analysis, we need to disentangle these
two contributions. The NMR $1/T_{1}T$ is given by:

\begin{equation}
\frac{1}{T_{1}T}=\gamma_{g}^{2}\lim_{\omega\rightarrow0}\sum_{\mathbf{k}}F^{2}\left(\mathbf{k}\right)\frac{\mathrm{Im}\,\chi\left(\mathbf{k},\omega\right)}{\omega}\label{T1T}
\end{equation}
where $\chi\left(\mathbf{k},\omega\right)$ denotes the dynamic magnetic
susceptibility at momentum $\mathbf{k}$ and frequency $\omega$.
Here, $\gamma_{g}$ is the constant gyromagnetic ratio and $F\left(\mathbf{k}\right)$
is the structure factor of the hyperfine interaction, which depends
on the direction of the applied field. The critical magnetic fluctuations
are associated with the ordering vectors $\mathbf{Q}_{1}=\left(\pi,0\right)$
or $\mathbf{Q}_{2}=\left(0,\pi\right)$ of the magnetic striped state,
and lead to the divergence of $1/T_{1}T$ as the temperature is lowered
towards the magnetic transition. By choosing the applied magnetic
field parallel to the FeAs plane, the structure factor $F\left(\mathbf{k}\right)$
is enhanced at the ordering vector $\mathbf{Q}$, favoring the dominant
contribution of the critical fluctuations \cite{Shannon11}. To remove
the non-critical Korringa\textbf{ }contribution coming from small
momenta $\mathbf{k}\approx0$, we follow Ref. \cite{Ning10} and subtract
from $1/T_{1}T$ the data of a heavily overdoped composition ($x=0.14$)
which is sufficiently far from magnetic and structural instabilities.
This is justified in this family of compounds due to the nearly doping-independent
shape of the Knight shift, which may be different in other series
\cite{NMR_Oka,NMR_Nakai,NMR_Hirano}. We note, however, that our scaling
analysis is robust and holds even for other choices of background
subtraction.\textbf{ }

Similarly, strongly overdoped samples display a rather different temperature
dependence for the shear modulus than underdoped and optimally doped
samples. While in the latter a critical softening of the shear modulus
is observed as the temperature decreases, in the former $C_{s}$ hardens
slightly at low temperatures because of phonon anharmonicity. Thus,
to obtain the critical contribution to the shear modulus, we used
the data of a heavily overdoped sample ($x=0.33$) as background,
as explained in \cite{Bohmer}. 

Figure \ref{fig_parent} presents both the shear modulus data of Ref.
\cite{Bohmer} (continuous curve) and the rescaled spin-lattice relaxation
rate (\ref{T1T}) of Ref. \cite{Ning10} (closed symbols) for the
undoped ($x=0$) composition. The agreement is excellent for the entire
temperature range, providing strong support for the existence of a
true scaling between spin and lattice fluctuations. Interestingly,
recent Raman scattering measurements in the tetragonal phase of the
same compound found that the orbital fluctuations are not strong enough
to account for the experimentally observed softening of the shear
modulus \cite{Gallais13}. For completeness, we also display in Fig.
\ref{fig_parent} the shear modulus data of Ref. \cite{Yoshizawa12}
(open symbols), to show the agreement between the ultrasound technique
of the latter work and the three-point bending method of Ref. \cite{Bohmer}.
We note that the differences in the two sets of data arise mostly
from disparities in the transition temperatures associated with sample
preparation. 

\begin{figure}
\begin{centering}
\includegraphics[width=0.85\columnwidth]{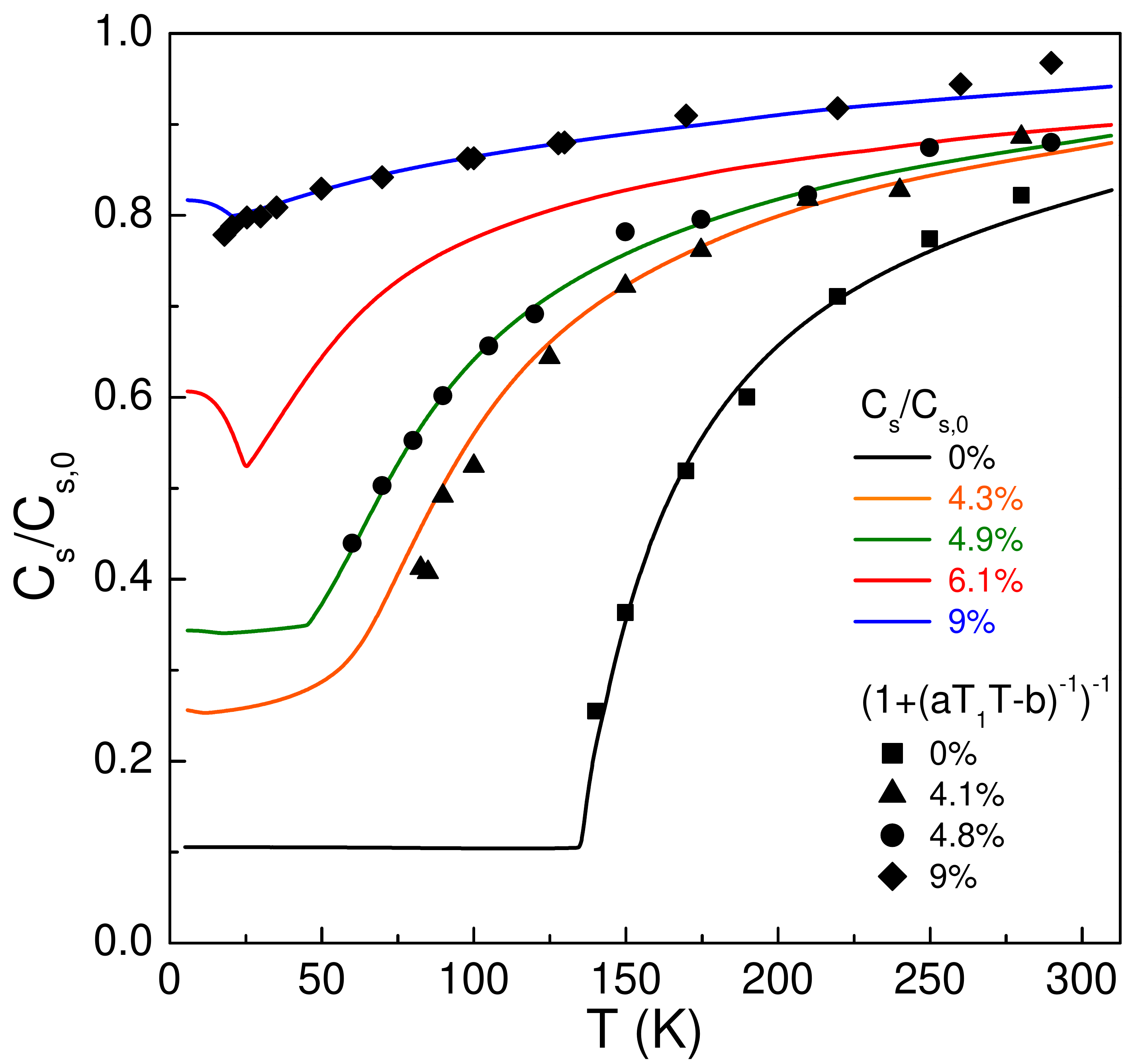} 
\par\end{centering}

\caption{Comparison between the relative shear modulus $C_{s}/C_{s,0}$ (continuous
lines, from Ref. \cite{Bohmer}) and the rescaled NMR $1/T_{1}T$
(closed symbols, from Ref. \cite{Ning10}) for different ``effective''
Co concentrations in $\mathrm{Ba(Fe_{1-x}Co_{x})_{2}As_{2}}$. The
``effective'' Co concentration was determined by comparing the available
transition temperatures ($T_{s}$, $T_{N}$, $T_{c}$) of the two
sets of samples with a third independent phase diagram (from Ref.\cite{Chu09}).}

\label{fig_doped} 
\end{figure}

To show that this agreement is not fortuitous, we also analyzed doped
samples, see Fig. \ref{fig_doped}. In this case, the comparison is
complicated by the fact that the two groups in Refs. \cite{Ning10}
and \cite{Bohmer} did not use the same samples and the determination
of the Co content may differ. For this reason, we determine an ``effective''
Co-content by comparing the available transition temperatures ($T_{s}$,
$T_{N}$, $T_{c}$) of the two sets of samples with a third independent
phase diagram (from Ref. \cite{Chu09}). The values are indicated
in Fig. \ref{fig_doped} and differ only slightly from those given
in the respective references. 

Our analysis of the rescaled $T_{1}T$ data presented in Fig. \ref{fig_doped}
remarkably captures the doping evolution of the temperature-dependent
shear modulus. The doping dependence of the scaling parameters $a$
and $b$ is shown in Fig. \ref{fig_phase_diagram}. While $a$ is
roughly constant across the entire phase diagram, $b$ is significantly
suppressed, approaching zero near optimal doping. Note from Eq. (\ref{scaling})
that $b$ is not a Curie-Weiss temperature, but a parameter that measures
the separation between the magnetic and structural instabilities.
Thus, its vanishing suggests that\textbf{ }the two instabilities tend
to the same temperature. We will come back to this point below.

Having established experimentally the existence of the scaling relation
(\ref{scaling}), we now discuss its origin using a general magneto-elastic
model \cite{FernandesPRL10,Cano10,Qi09,Gorkov09,Indranil11}. Since
we removed the non-critical contributions to $1/T_{1}T$ and $C_{s}$,
a low-energy model suffices. In particular, we consider two magnetic
order parameters $\mathbf{M}_{1}$ and $\mathbf{M}_{2}$, referring
to the magnetic striped states with ordering vectors $\mathbf{Q}_{1}=\left(\pi,0\right)$
(i.e. spins parallel along $\hat{\mathbf{y}}$ and anti-parallel along
$\hat{\mathbf{x}}$) and $\mathbf{Q}_{2}=\left(0,\pi\right)$ (i.e.
spins parallel along $\hat{\mathbf{x}}$ and anti-parallel along $\hat{\mathbf{y}}$)
\cite{Fernandes12}. We also include the orthorhombic order parameter
$\delta=\epsilon_{xx}-\epsilon_{yy}$, where $\epsilon_{ij}\equiv\frac{1}{2}\left(\partial_{i}u_{j}+\partial_{j}u_{i}\right)$
is the strain tensor. For simplicity, we consider here the coordinate
system referring to the 1-Fe unit cell. The action of the collective
magnetic degrees of freedom is given by: 

\begin{align}
S_{\mathrm{mag}}= & \sum_{q,i}\chi^{-1}\left(q\right)\mathbf{M}_{i,q}\cdot\mathbf{M}_{i,-q}+\frac{u}{2}\sum_{r}\left(M_{1}^{2}+M_{2}^{2}\right)^{2}\notag\\
 & -\frac{g_{0}}{2}\sum_{r}\left(M_{1}^{2}-M_{2}^{2}\right)^{2}\label{F_mag}
\end{align}
where $u,g_{0}>0$ are constants and we introduced the notations $q=\left(\mathbf{q},\omega\rightarrow i\omega_{n}-i0^{+}\right)$
and $r=\left(\mathbf{r},\tau\right)$, with $\tau$ denoting imaginary
time and $\omega_{n}=2n\pi T$, Matsubara frequency. At the mean-field
level, minimization of this free energy leads to the two magnetic
stripe configurations when $\chi^{-1}\left(\mathbf{Q}\right)\rightarrow0$,
i.e. we obtain either $M_{1}=0$, $M_{2}\neq0$ or $M_{1}\neq0$,
$M_{2}=0$. This free energy can in fact be microscopically derived
from models of either itinerant electrons \cite{Fernandes12,Brydon11,Haas13}
or localized spins \cite{Fang08,batista}.

The lattice contribution to the action is given by:

\begin{equation}
S_{\mathrm{el}}=\sum_{r}\left[\frac{1}{2}C_{s,0}\delta^{2}-\lambda\delta\left(M_{1}^{2}-M_{2}^{2}\right)\right]\label{F_el}
\end{equation}

The first term is just the harmonic part of the elastic energy, while
the second one is the magneto-elastic coupling, with arbitrary coupling
constant $\lambda>0$. This coupling makes bonds connecting anti-parallel
(parallel) spins expand (shrink) in the orthorhombic phase (see inset
of Fig. \ref{fig_parent}). In principle, one could assume that $C_{s,0}$
itself becomes zero at a certain temperature. Here, instead, we assume
$C_{s,0}$ to be constant and to never become soft on its own. Then
we can integrate out the Gaussian orthorhombic fluctuations and derive
the shear modulus $C_{s}$ renormalized by spin fluctuations, obtaining
\textbf{$\chi_{\mathrm{nem}}$} of Eq.\ref{chinem} as \textbf{$\chi_{\mathrm{nem}}^{-1}=\chi_{0,\mathrm{nem}}^{-1}-g$}.
The bare nematic susceptibility is determined by the dynamic spin-susceptibility
\textbf{$\chi\left(q\right)$} properly renormalized by magnetic fluctuations,
$\chi_{0,\mathrm{nem}}=\sum\limits _{q}\chi^{2}\left(q\right)$. The
nematic coupling $g$ is the bare coupling $g_{0}$ of Eq. (\ref{F_mag})
renormalized by all other non-soft modes, such as elastic fluctuations
and the ferro-orbital susceptibility $\chi_{\mathrm{orb}}$, which
leads to\textbf{ }an enhancement\textbf{ }$g\rightarrow g+\lambda_{\mathrm{orb}}^{2}\chi_{\mathrm{orb}}$
\cite{SUST12}. 

To make contact with the spin-lattice relaxation rate (\ref{T1T}),
we note that the magnetic susceptibility has overdamped dynamics near
the ordering vectors $\mathbf{Q}$, i.e. $\chi^{-1}\left(\mathbf{q},\omega\right)=\chi^{-1}\left(\mathbf{q}\right)-i\omega\Gamma$,
with Landau damping $\Gamma$. Substituting in Eq. (\ref{T1T}), we
obtain $1/T_{1}T=\gamma_{\mathrm{g}}^{2}\Gamma F^{2}\left(\mathbf{Q}\right)\sum\limits _{\mathbf{q}}\chi^{2}\left(\mathbf{q}\right)$,
where we replaced $F\left(\mathbf{q}\right)\rightarrow F\left(\mathbf{Q}\right)$
because of the direction of the applied field. Now, going back to
the spin-nematic expression for $\chi_{0,\mathrm{nem}}$, we note
that if the system is in the vicinity of a finite-temperature critical
point - which is the certainly the case for underdoped samples - we
can replace the sum over momentum and Matsubara frequency by a sum
over momentum only, i.e. $\sum\limits _{q}\chi^{2}\left(q\right)\rightarrow T_{0}\sum\limits _{\mathbf{q}}\chi^{2}\left(\mathbf{q}\right)$,
where $T_{0}$ is the temperature scale associated with the magnetic
transition. Then, using the expression for $1/T_{1}T$, we obtain
the scaling (\ref{scaling}) with:

\begin{equation}
a=\frac{\gamma_{\mathrm{g}}^{2}\Gamma F^{2}\left(\mathbf{Q}\right)C_{s,0}}{\lambda^{2}T_{0}}\:;\quad b=\frac{gC_{s,0}}{\lambda^{2}}\label{aux_Cs_theory}
\end{equation}

Since we assumed that magnetic fluctuations are the only soft mode,
this result confirms that the scaling (\ref{scaling}) is a signature
of a magnetically-driven structural transition, as we argued qualitatively
above. Note that a distinct possible relation between spin fluctuations
and elastic softening, $C_{s}=a-b/T_{1}T$, was mentioned in Ref.
\cite{Nakai13}. Our scaling relation in Eq. (\ref{scaling}), however,
not only connects both quantities in a quantitatively more accurate
way but it also has a very clear physical interpretation, as shown
above. 

Equation (\ref{aux_Cs_theory}) allows us to understand the doping
evolution of the parameter $b$ in Fig. \ref{fig_phase_diagram} from
a more physical perspective. As doping increases, we notice that this
parameter changes from $b>1$ in the undoped compound to $b\rightarrow0$
near optimal doping. We interpret this decrease of $b$ as an indication
that the system crosses over from a regime where the nematic coupling
is dominated by the magnetic contribution, $g>\lambda^{2}/C_{s,0}$,
to a regime governed by the magneto-elastic coupling, $g<\lambda^{2}/C_{s,0}$.
Since the bare high-temperature shear modulus barely changes with
doping (see Ref. \cite{Yoshizawa12}), either the magneto-elastic
coupling $\lambda$ increases or the bare coupling $g_{0}$ decreases
towards optimal doping. Interestingly, calculations of $g_{0}$ based
on itinerant electrons found a suppression of this quantity as charge
carriers are introduced \cite{Fernandes12}.

The tendency of a vanishing $b$ near optimal doping suggests that
the magnetic and elastic instabilities converge to the same point
\cite{Birgeneau09}. Indeed, by fitting the $T_{1}T$ and $C_{s}$
data with Curie-Weiss expressions $\left(T-T_{T_{1}T}\right)$ and
$\left(T-T_{C_{s}}\right)/\left(T-\theta_{C_{s}}\right)$, respectively,
we also find that the estimated bare transition temperatures $T_{T_{1}T}$
and $T_{C_{s}}$ tend to converge at optimal doping (see Fig. \ref{fig_phase_diagram}).
The negative value of $b$ in the slightly overdoped sample can have
different origins. One possible reason is the inadequacy of the above
derivation for the scaling relation near a quantum critical point,
where the Matsubara frequency becomes a continuous variable and $\sum\limits _{q}\chi^{2}\left(q\right)\neq T_{0}\sum\limits _{\mathbf{q}}\chi^{2}\left(\mathbf{q}\right)$.
A more interesting possibility is that $g$ itself becomes negative
in Eq. (\ref{aux_Cs_theory}). This would indicate that the magnetic
instability is not towards a striped magnetic state, but an SDW phase
that preserves the tetragonal symmetry of the system \cite{Eremin10}.
Interestingly, such a state has been recently found experimentally
in optimally hole-doped $\mathrm{(Ba_{1-x}Na_{x})Fe_{2}As_{2}}$ \cite{Osborn13}
and in $\mathrm{Ba}\mathrm{(Fe_{1-x}Mn_{x})_{2}As_{2}}$ \cite{Kim10}.

In summary, our analysis reveals a robust scaling relation between
the shear modulus and the NMR spin-lattice\textbf{ }relaxation rate\textbf{
}in\textbf{ }$\mathrm{Ba(Fe_{1-x}Co_{x})_{2}As_{2}}$. This result
unveils the fact that the ubiquituous elastic softening in\textbf{
}these iron pnictides is a consequence of the magnetic fluctuations
associated with the degenerate (i.e. frustrated) ground states with
ordering vectors $\mathbf{Q}_{1}=\left(\pi,0\right)$ and $\mathbf{Q}_{2}=\left(0,\pi\right)$.\textbf{
}Due to the similarity between the phase diagrams of $\mathrm{Ba(Fe_{1-x}Co_{x})_{2}As_{2}}$
and of other iron-pnictide families, we expect this scaling relationship
to hold in other compounds as well.

We thank A. Chubukov, P. Chandra, V. Keppens, D. Mandrus, and M. Yoshizawa
for fruitful discussions. We are grateful to Y. Matsuda for pointing
out Ref. \cite{Nakai13} to us. A part of this work was supported
by the DFG under the priority program SPP1458.

\end{document}